\documentstyle[epsfig,12pt]{elsart}

\begin{document}

\begin{frontmatter}
\title{Stopping and Isospin Equilibration in Heavy Ion Collisions}

\author[catania]{T. Gaitanos},
\author[catania]{M. Colonna},
\author[catania]{M. Di Toro\thanksref{dit}},
\author[muenchen]{H.~H. Wolter}
\thanks[dit]{ditoro@lns.infn.it}
\address[catania]{Laboratori Nazionali del Sud INFN, I-95123 Catania, Italy
\\ Physics and Astronomy Dept., Univ. of Catania  } 
\address[muenchen]{Sektion Physik, Universit\"at M\"unchen, 
D-85748 Garching, Germany}  

\begin{abstract}
We investigate the density behaviour of the symmetry energy 
with respect to isospin equilibration in the combined systems 
$Ru(Zr)+Zr(Ru)$ at relativistic energies of $0.4$ and $1.528~AGeV$. 
The study 
is performed within a 
relativistic framework and the contribution 
of the iso-vector, scalar $\delta$ field to 
the symmetry energy and the isospin dynamics is particularly explored. 
We find that the isospin mixing depends on the symmetry energy
and a stiffer behaviour leads to more transparency.
The results are also nicely sensitive to the ``fine structure'' of 
the symmetry energy, i.e. to the covariant properties of 
the isovector meson fields.

The isospin tracing appears much less dependent on the in medium
neutron-proton cross sections ($\sigma_{np}$) and this makes
such observable very peculiar for the study of the
isovector part of the nuclear equation of state.
 
Within such a framework, 
comparisons with experiments support the introduction of the 
$\delta$ meson in the description of the iso-vector equation of state.
\end{abstract}

\end{frontmatter}
The particular motivation for high energy heavy ion collisions has been 
to determine 
the nuclear equation-of-state ($EOS$) at densities away from 
saturation and at non-zero 
temperatures \cite{fopir}. Recently the isospin degree of freedom, i.e. the 
$EOS$ of asymmetric nuclear matter, has become of particular 
interest \cite{iso}. 
The asymmetry term at normal nuclear density is relatively 
well known from the mass 
formula and its momentum dependence from optical potential fits (Lane 
potential, ref. \cite{lane}). 
However, the iso-vector 
$EOS$ is empirically poorly determined for densities beyond saturation 
and actually it is very 
differently predicted from the various nuclear many-body models \cite{iso1}. 
On the other hand, 
the iso-vector $EOS$ is of crucial importance in extrapolating 
structure calculations 
away from the valley of stability and in astrophysical processes, 
such as neutron star cooling and supernova explosions. 

So far asymmetric nuclear matter has been only poorly studied 
for extreme conditions 
beyond saturation. Studies on finite nuclei \cite{rmf} refer only to 
ground state 
nuclear matter, while at supra-normal densities one has to relay 
on extrapolations. Thus, more useful information could be achieved in 
studies of 
heavy ion collisions due to the formation of hot and 
dense asymmetric matter for short time scales during the reaction dynamics. 
Theoretical studies of asymmetric nuclear matter beyond 
saturation have been recently started. They can be divided into two groups, 
non-relativistic using phenomenological Skyrme-interactions 
\cite{bao03} and 
relativistic ones based on the Relativistic Mean Field (RMF) theory of the 
Quantum Hadrodynamics ($QHD$) \cite{qhd,liu,grefl,gait04}. 

In this study we use a common relativistic framework for the static 
description of 
asymmetric nuclear matter \cite{liu} and for the dynamic case of 
heavy ion collisions \cite{gait04}. 
This is particularly appropriate for heavy ion collisions at intermediate 
energies, since such a covariant formulation intrinsically 
predicts the main features 
of the density and momentum dependence of the nuclear $EOS$ due to 
the relative importance of the scalar and vector fields
in the reaction dynamics. The relativistic structure is of great 
interest in the iso-vector sector for the competition between 
Lorentz vector ($\rho$) and 
Lorentz scalar ($\delta$) iso-vector fields, which show a similar 
cancellation and similar 
dynamic effects as in the iso-scalar ($\sigma, \omega$) sector. 
While structure calculations 
seem to be rather insensitive to the presence of a $\delta$ field, 
we expect clearly distinguishable effects at higher densities. 
Indeed the high density 
dependence of the symmetry energy $E_{sym}$ is particularly affected by 
the different treatment of the 
microscopic Lorentz structure of the iso-vector part 
of the nuclear mean field \cite{liu,grefl,gait04}. 
 
$E_{sym}$ becomes stiffer due to the introduction of the attractive
scalar $\delta$ field, 
for a pure relativistic mechanism.
The iso-vector $EOS$ can be characterized by a vector $\rho$ meson or, 
as in the iso-scalar case, by a competition of vector $\rho$ and 
scalar $\delta$ field contributions.
As discussed in detail in Refs. \cite{liu,grelin}, with the same 
fixed bulk asymmetry parameter in 
both cases, one has to increase the $\rho$ meson coupling in the presence of 
the attractive 
$\delta$ field. On the other hand, the latter field, due to its Lorentz scalar 
character, is proportional to an invariant scalar density 
$\rho_{s} \approx \frac{m^{*}}{E_{F}^{*}}\rho_{B}$, suppressed for 
baryon densities 
above saturation $\rho_{B} >> \rho_{sat}$. Thus, this 
competition between Lorentz 
vector ($\rho$) and Lorentz scalar ($\delta$) iso-vector fields 
leads finally to a 
stiffer behavior of the symmetry energy at high densities.
The scalar nature of the $\delta$-field will also naturally imply
a definite neutron/proton effective mass splitting \cite{liu,grelin},
with a more reduced neutron mass at high density. 

The consideration of the $\delta$ meson in the description of 
asymmetric nuclear matter gives then rise to a larger repulsion 
(attraction) for neutrons (protons) for 
baryon densities $\rho_{B}$ above saturation. 
Moreover the larger $\rho$-meson coupling is further increasing the
repulsion seen by high momentum neutrons due to the Lorentz boosting
of the charged vector field, as already observed in flow studies,
see the discussion in ref.\cite{grefl}. 
These effects together 
with the 
effective mass splitting are responsible for new interesting 
transport phenomena which 
could help to determine the high density behavior of the symmetry energy
and its microscopic structure. 

\begin{figure}
\begin{center}
\includegraphics[scale=0.35]{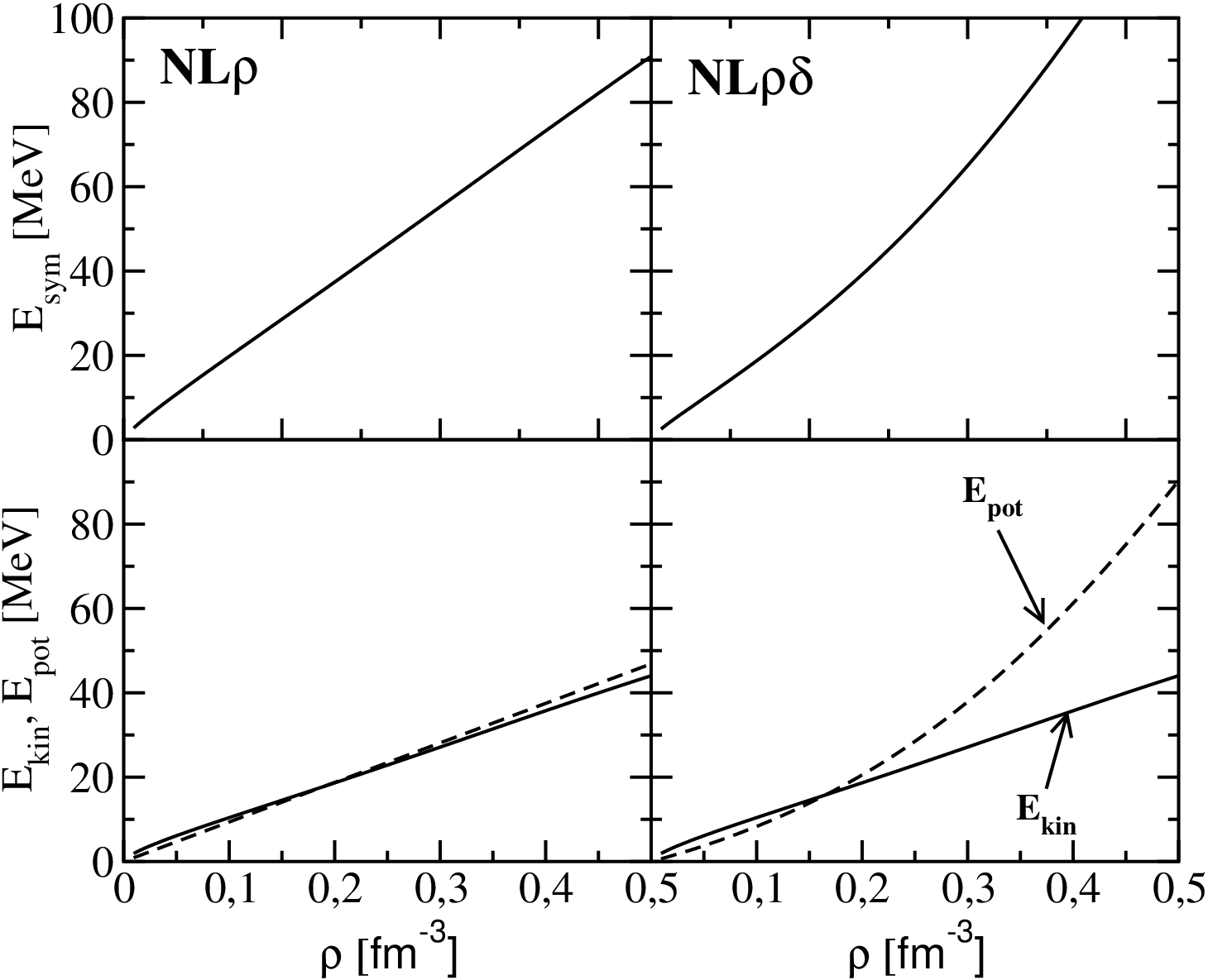}
\caption{\label{Fig1} 
(Top) Density dependence of the symmetry energy $E_{sym}$ for the Non-Linear 
$Walecka$-model ($NL$) including only the $\rho$ field ($NL\rho$) and both, 
the $\rho$ and $\delta$ fields ($NL\rho\delta$) in the iso-vector sector. 
The pannels in the bottom show the kinetic ($E_{kin}$) and potential 
($E_{pot}$) contributions to the total symmetry energy $E_{sym}$ separately.
}
\end{center}
\end{figure}

The relativistic features of the iso-vector sector are shown 
in Fig. \ref{Fig1} in terms 
of the symmetry energy $E_{sym}$ within 
the non-linear $RMF$ model 
including only 
the iso-vector, vector $\rho$ ($NL\rho$) and both, the iso-vector, 
vector $\rho$ and 
iso-vector, scalar $\delta$ mesons ($NL\rho\delta$).
The choice of the corresponding coupling constants is discussed in
detail in the refs. \cite{grefl,gait04}. The related kinetic equations are
used for transport simulations of the reaction dynamics of the
relativistic heavy ion collisions presented here  \cite{gait04}.
We solve the covariant transport equation of the Boltzmann type
within the Relativistic Landau Vlasov ($RLV$) method \cite{fw95} 
(for the Vlasov part) and applying a Monte-Carlo procedure for the collision
term, including inelastic processes involving the production/absorption
of nucleon resonances \cite{ha94}.

In order to reach definitive conclusions on the high density behavior 
of the symmetry energy one 
should first select the key signals and then simultaneously compare 
with experimental 
data. Collective isospin flows \cite{grefl,bao1} and particle production, 
i.e. pion \cite{bao03,gait04,bao1} and 
kaon spectra \cite{ferini} seem to be good candidates. Here we continue 
the previous studies 
by considering another interesting aspect, which has been 
extensively investigated by 
experiments of the $FOPI$ collaboration: the correlation between the degree 
of stopping and the isospin equilibration (or transparency). 

The idea is to study 
colliding systems with the same mass number but different $N/Z$ ratio, 
in particular 
a combination of ${}^{96}_{44}Ru,~N/Z=1.18$ and ${}^{96}_{40}Zr,~N/Z=1.4$ 
has been 
used as projectile/target in experiments at intermediate energies of $0.4$ and 
$1.528~AGeV$ \cite{fopi3,fopi4}. 
The degree of stopping or transparency has been determined by 
studying the rapidity dependence of the {\it imbalance ratio} for the 
mixed reactions 
$Ru(Zr)+Zr(Ru)$: $R(y^{(0)})=N^{RuZr}(y^{(0)})/N^{ZrRu}(y^{(0)})$, 
where $N^{i}(y^{(0)})$ is the particle emission yield inside the detector 
acceptance at a given rapidity for $Ru+Zr,~Zr+Ru$ with $i=RuZr,~ZrRu$. The 
observable $R$ can be measured for different particle species, 
like protons, neutrons, light fragments such as $t$ and ${}^{3}He$ and 
produced particles such as pions ($\pi^{0,\pm}$), etc. It characterizes 
different stopping scenarios.

In the proton case, moving from
target to $cm$ rapidity, $R(p)$ rises 
(positive slope) 
for partial transparency, 
falls (negative slope) for full stopping/rebound scenarios and it is flat 
when total isospin mixing is achieved in the collision.
An opposite behavior will appear for neutrons. 
Indeed we remind that in the full transparency limit $R$ should 
approach the initial value of 
$R(p)=Z^{Zr}/Z^{Ru}=40/44=0.91$ and $R(n)=N^{Zr}/N^{Ru}=56/52=1.077$ for 
protons and 
neutrons at target rapidity, respectively. 
Therefore, $R(p)$ can be regarded as a sensitive 
observable with respect to isospin diffusion, i.e. to properties of the 
symmetry term as we will show later.

In order to proceed to a direct comparison with experiments, in our
kinetic equation simulations we have used a coalescence method to
select nucleon and light ion emissions. The algorithm is quite
standard. As described in ref.\cite{gait01}, the coalescence radii
in coordinate ($<R>_c = 4 fm$) and in
momentum ($<R>_p = 1.3 fm^{-1}$) space are fixed from the
experimental charge distributions.
In the present calculations we have also checked the
stability of the yields by varying the coalescence parameters
($<R>_{c,p}$) up to a $20\%$. 

For the investigation of isospin equilibration  we have 
analyzed the transport results in the same way as carried out in the 
$FOPI$ experiment, 
in particular selecting central events through the observable $ERAT$ 
which measures the ratio of the mean transverse to the mean longitudinal 
kinetic energy. To do so, we first apply the coalescence method
for fragment 
production in the final state and then we calculate the observable $ERAT$. 
A detailed 
description of this analysis can be found in ref. \cite{gait01}, 
where it was shown 
that the  charged particle multiplicity 
and $ERAT$ distributions fit well the 
experimental data, as an important check of the phenomenological 
phase space model. At this point we can use the same centrality cuts
as in the $FOPI$ exps. \cite{fopi4}.

\begin{figure}
\begin{center}
\includegraphics[scale=0.35]{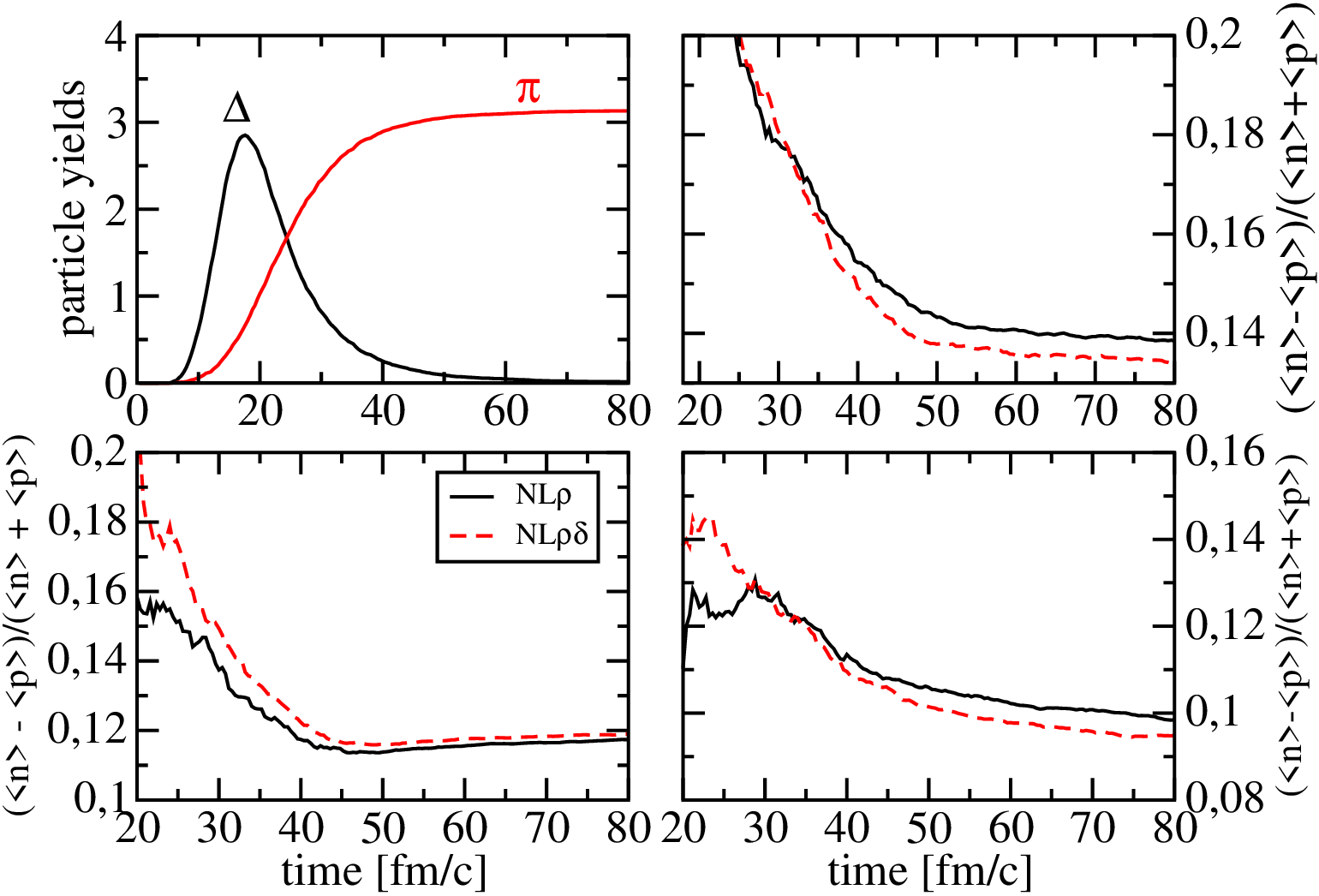}
\caption{Central ($b<2~fm$) $Ru+Zr$ 
collisions at $0.4~AGeV$ beam energy. Top-left: time evolution of
 $\Delta$-resonance and $\pi$ production rate.
Time evolution of the asymmetry parameter $\frac{<n>-<p>}{<n>-<p>}$ for 
the emitted particles: (top-right) for particles at target 
rapidity, (bottom-left) at mid rapidity and (bottom right) at the projectile 
rapidity for the same models of the previous figure. 
$NL\rho$: solid lines; $NL\rho\delta$: dashed lines. }
\label{Fig2}
\end{center}
\end{figure}

We start the discussion of the results showing, in the case of a central
$Ru+Zr$ collision at $=0.4AGeV$, the time evolution of 
the asymmetry 
parameter $\frac{<n>-<p>}{<n>+<p>}$, with $<\cdots>$ being the average 
number of 
emitted particles (protons, neutrons), for different rapidity regions as 
indicated 
(see Fig. \ref{Fig2}). The compression/expansion time scales are given 
in terms 
of the temporal development of the $\Delta$ resonances and the produced 
pions ($\pi$). 

We see that the colliding system emits more neutrons (increase of the
asymmetry parameter) in the transport 
calculations with the $NL\rho\delta$ interaction. The effect is more 
pronounced during the compression
 phase ($15 < t < 30~fm/c$), 
in particular in the mid-rapidity region, whereas in the 
spectator regions the trend becomes opposite for late time scales. 

The observed differences between $NL\rho$ and $NL\rho\delta$ models can 
be understood 
in terms of the density dependence of the symmetry energy, Fig. \ref{Fig1}. 
Its stiff character in the $NL\rho\delta$ model generates stronger (smaller) 
pressure gradients being responsible for more repulsion (attraction) 
for neutrons (protons). This is also consistent with the effective mass 
splitting, where the 
neutrons (protons) experiences a smaller (larger)  effective mass for 
densities beyond saturation. Thus, neutrons are emitted earlier than protons 
increasing the asymmetry of the emitted particles. The isospin effects of 
the $\delta$ meson are more pronounced only during and just after 
the compression 
phase, since the different treatment of the microscopic Lorentz 
structure of the 
symmetry energy affects it only for supra-normal densities, 
see also Fig. \ref{Fig1}.  Moreover for high momentum particles we have
a pure relativistic boosting of the vector $\rho$-meson repulsion for
neutrons, and attraction for protons, see the discussion of ref.\cite{grefl}.
 
The opposite late trends (expansion phase) in the asymmetry parameter of 
emitted particles 
at projectile/target 
rapidity can be consistently accounted for by the low density behaviour 
of the symmetry energy. In this 
region the $\delta$-coupling in neutron-rich matter leads to 
more (less) attraction for 
neutrons (protons) responsible for a reduction of neutron emission. 
However, it is 
rather a moderate effect as expected from the low density 
dependence of the symmetry energy shown in Fig. \ref{Fig1}.

\begin{figure*}
\begin{center}
\includegraphics[scale=0.50]{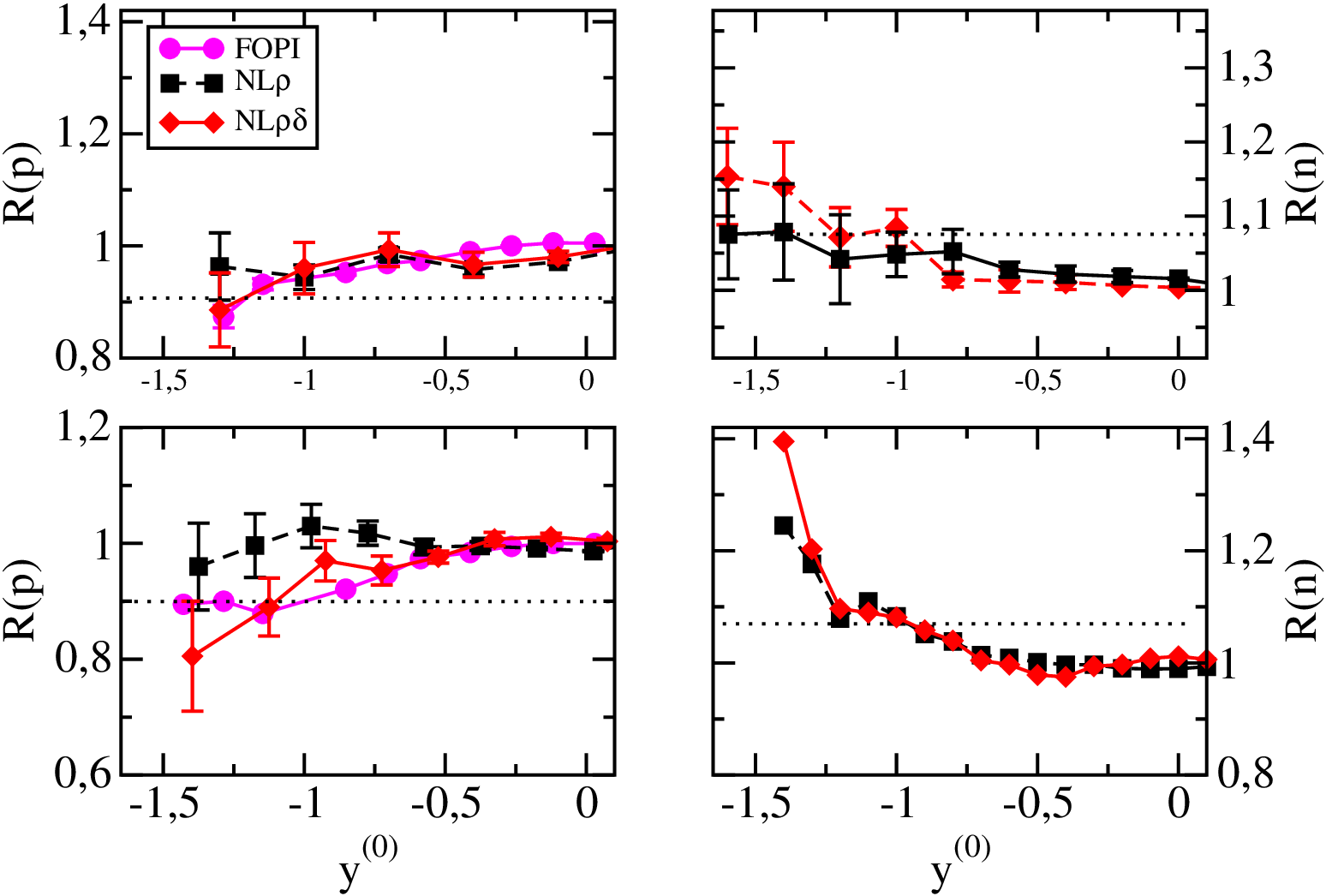}
\caption{\label{Fig3} 
The imbalance ratio $R(y^{(0)})=\frac{N^{RuZr}(y^{(0)})}{N^{ZrRu}(y^{(0)})}$ 
as 
function of the normalized rapidity $y^{(0)}=y/y_{proj}$ of free protons 
(left columns) and free neutrons (right columns) for the models of Fig. 
\protect\ref{Fig1}, for central ($b\leq 2~fm$) $Ru(Zr)+Zr(Ru)$-collisions 
at $0.4$ (top) and $1.528~AGeV$ beam energies. 
The experimental data are taken from the $FOPI$ collaboration 
\protect\cite{fopi4}.
}
\end{center}
\end{figure*}

\begin{figure*}
\begin{center}
\includegraphics[scale=0.40]{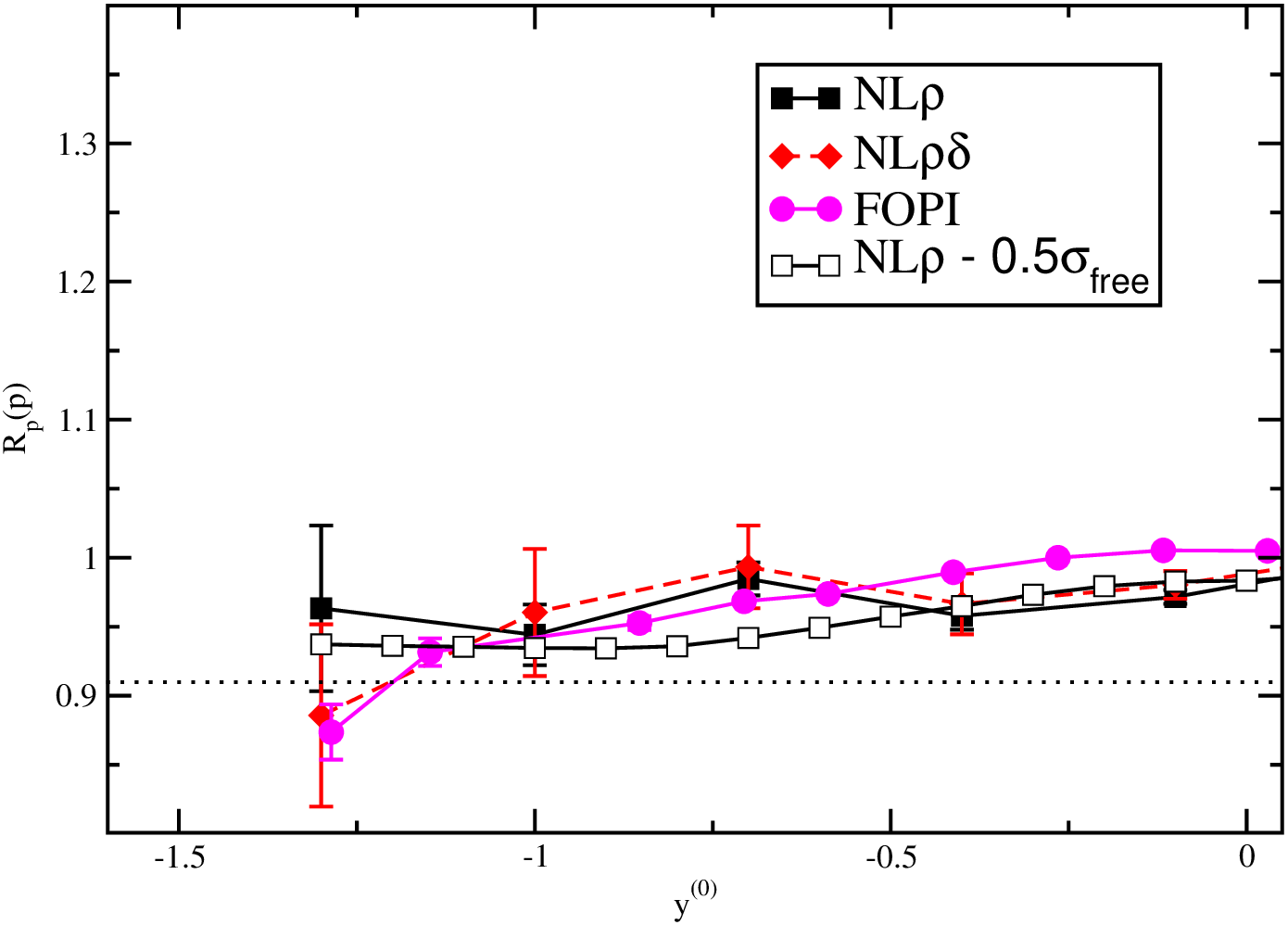}
\caption{\label{Fig4} 
Proton imbalance ratio as 
function of the normalized rapidity for central ($b\leq 2~fm$) 
$Ru(Zr)+Zr(Ru)$-collisions 
at $0.4~AGeV$ like the top-left part of Fig.\ref{Fig3}.
A curve is added (empty squares) corresponding to a $NL\rho$ calculation
with reduced nucleon-nucleon cross sections (half the free values,
 $\sigma_{NN}(E) = \frac{1}{2} \sigma_{free}$).  
}
\end{center}
\end{figure*}

In Fig. \ref{Fig3}, we report the rapidity dependence of 
the imbalance ratio in central collisions
of the mixed system $Ru(Zr)+Zr(Ru)$ for free protons ($R(p)$) and 
free neutrons ($R(n)$) at the two energies
$0.4$ and $1.528~AGeV$. 

First of all, the imbalance ratio correctly approaches 
unity at mid-rapidity for all particle types considered.
This is a obvious check of the calculation since the symmetry of the two 
collisions naturally implies a full 
isospin mixing in the $c.m.$ rapidity region.
 
Going from target- to mid-rapidity it nicely rises for protons,
 and decreases for neutrons, a good isospin transparency signature. 
The effect is much more evident for the $NL\rho\delta$ interaction.
 The observed difference between 
the two models is obvious since within the $NL\rho\delta$ picture 
neutrons experience
a more repulsive iso-vector mean field, particularly at high densities, than 
 protons, with a consequent much less nucleon stopping in the 
colliding system. 
 This picture is consistent with the 
previous Fig. \ref{Fig2}. It can be also considered as a high density
 effect since the 
particle emission takes mainly place during and just after 
the compression phase. 

The observed effect of the $EOS$ on the imbalance ratio of protons 
and neutrons is 
not very large. At low intermediate energies ($0.4~AGeV$) 
one has to deal with moderate compressions of $\rho_{B} < 2 \cdot \rho_{sat}$ 
where the differences in the iso-vector $EOS$ arising from 
the $\delta$ meson and thus from 
the different treatment of the Lorentz structure are small. 
On the other hand, at higher 
incident energies ($1.528~AGeV$) 
a larger difference is observed between the two models. 
Indeed, one observes a slightly higher 
isospin effect on the imbalance ratios. However, 
with increasing beam energy the 
opening of inelastic channels via the 
production/decay of $\Delta$ resonances through pions and 
additional secondary scattering 
(pion absorption) with isospin exchange, e.g. 
$nn \longrightarrow n\Delta^{0},~\Delta^{0} \longrightarrow p\pi^{-}$, 
also contributes as a background effect to the final result. 
This interpretation is 
confirmed by other studies \cite{bao03}. 

Reduced in medium Nucleon-Nucleon ($NN$) cross sections, in particular the
$\sigma_{np}$, will also increase the isospin transparency in the direction
shown by the data. We have tested this possibility considering a factor
two reduction, $\sigma = \frac{1}{2} \sigma_{free}$, of the free 
 $NN$-cross section values used before. We note that such a reduction
represents a rather strong in-medium effects as compared to
recent microscopic Dirac-Brueckner estimations \cite{fuch}.

In Fig.\ref{Fig4} we report the results, open squares, for the proton 
imbalance ratios at $0.4~AGeV~+~0.5\sigma_{free}$ in the $NL\rho$
case. We see a overall slightly increasing transparency but not enough
to reproduce the trend of the experimental values in the target rapidity 
region. On the other hand the reduction of the $NN$ cross sections,
and in particular of the $\sigma_{(np)}$, implies a too large
transparency in the proton rapidity distributions for central collisions
of the charge symmetric $Ru~+~Ru$ case; effect clearly seen in our
simulations and already remarked in previous $IQMD$ calculations,
 see ref.\cite{fopi4}. So we can exclude such and further reductions
of the $NN$ cross sections.

Since the calculations are performed with the same $EOS$ for the
symmetric nuclear matter, same compressibility and momentum dependence
 \cite{comp}, the observed transparency appears to be uniquely
related to $isovector-EOS$ effects, i.e. to the isospin dependence
of the nucleon self-energies at high baryon densities.

\begin{figure}[t]
\unitlength1cm
\begin{picture}(8.,7.0)
\put(0.0,0.0){\makebox{\epsfig{file=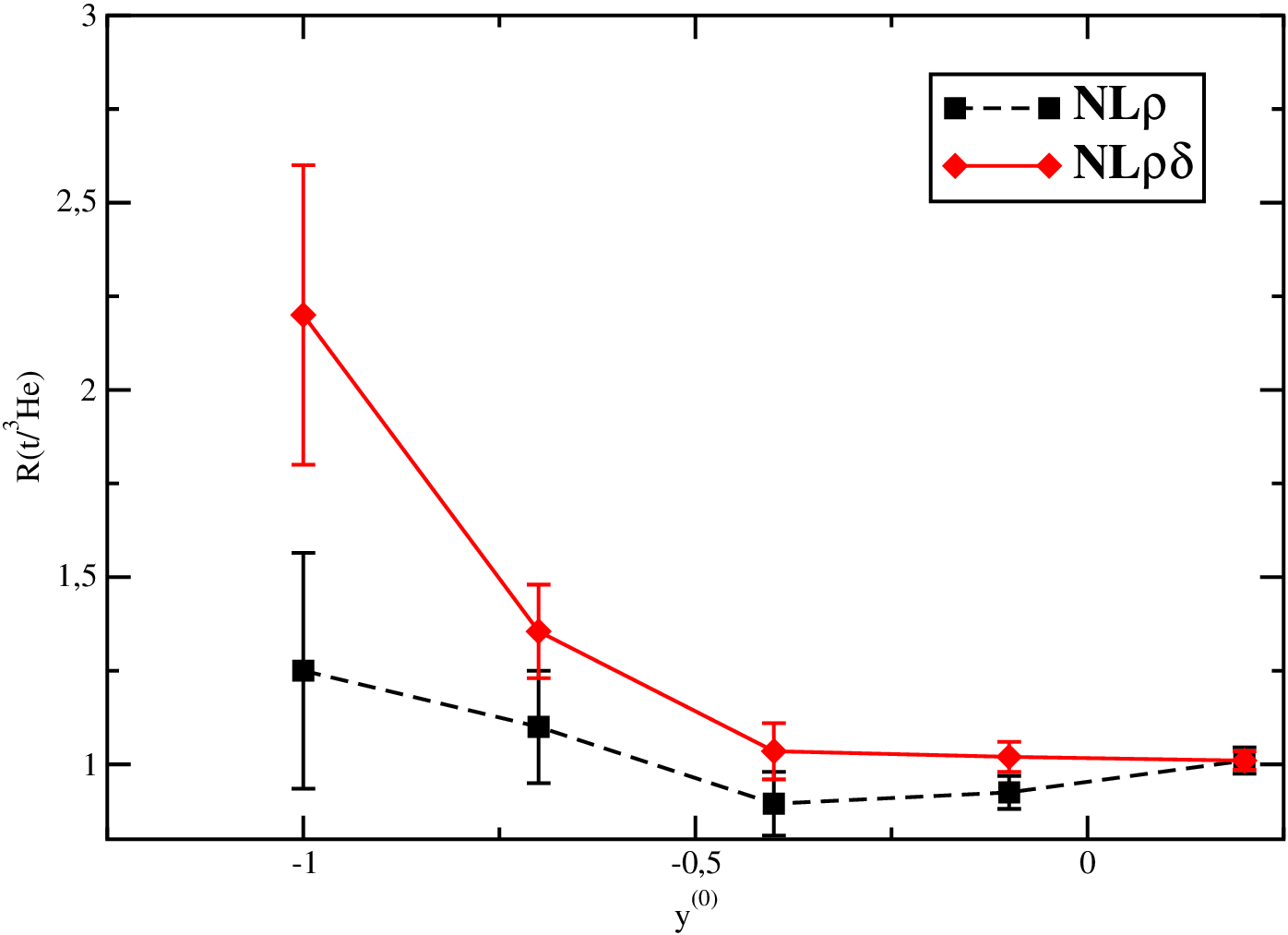,width=6.9cm}}}
\put(7.2,0.0){\makebox{\epsfig{file=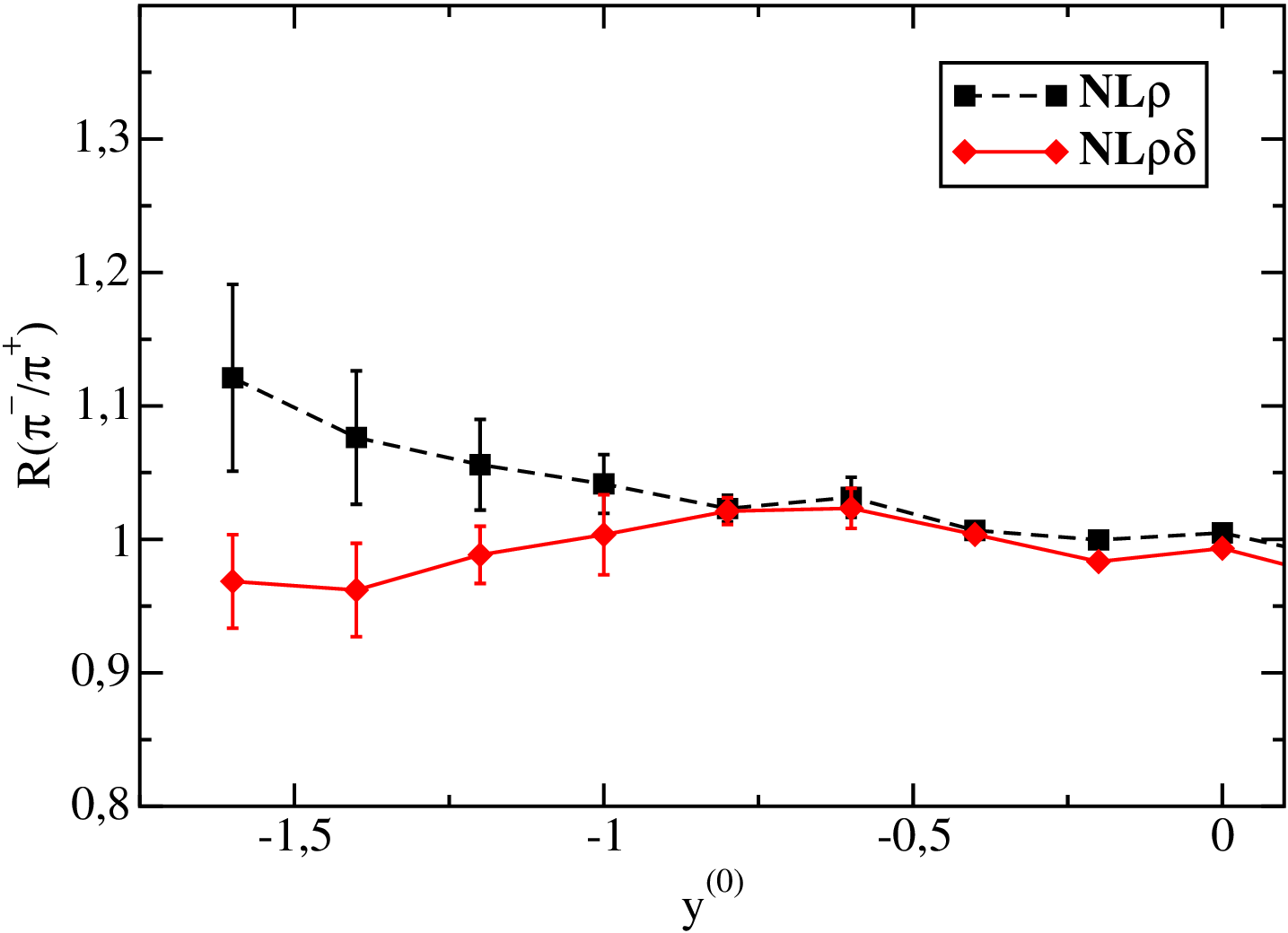,width=6.9cm}}}
\end{picture}
\caption{Same as in Fig.~\ref{Fig3} for $0.4~AGeV$ beam energy, but 
for the ratio of $t$ to ${}^{3}He$ (left) and the ratio of negative 
to positive charged pions (right).
}
\label{Fig5}
\end{figure}


In Fig.\ref{Fig5} we report the rapidity dependence of the imbalance
ratio for other particle emissions in central collisions
of the mixed system $Ru(Zr)+Zr(Ru)$ at $0.4~AGeV$ beam energy: (left) 
for the ratio of triton to $He$-fragments ($R(t/{}^{3}He)$) and (right) 
for the ratio of negative $\pi^{-}$ to positive $\pi^{+}$ charged pions 
($R(\pi^{-}/\pi^{+})$).

The fact that protons {\it and} neutrons
exhibit an {\it opposite behavior} for the corresponding imbalance 
ratios at target 
rapidity, clearly suggests that the detection of the imbalance observable 
$R(t/{}^{3}He)$ for the $t/{}^{3}He$ ratio should reveal a larger sensitivity. 
We thus have determined the observable $R(t/{}^{3}He)$, using
the above discussed phase space coalescence procedure to extract
the light ion emissions. 

Fig. \ref{Fig5} (on the left) shows the $R(t/{}^{3}He)$
results. The isospin effect 
originating from the 
appearence of the iso-vector, scalar $\delta$ meson in the 
$NL\rho\delta$ model 
turns out to be here crucial near target rapidities. We note that the 
effect indeed can be 
hardly seen from the separate imbalance ratios for protons and neutrons 
at the same 
rapidity, see Fig. \ref{Fig3} top panels, apart the difficulties 
of neutron emission detections. 
It would be therefore of great interest to experimentally measure 
directly this quantity.

Finally, another sensitive observable appears to be the imbalance 
ratio of charged pions 
$R(\pi^{-}/\pi^{+})$ (Fig. \ref{Fig5}, right panel). 
At variance with the previous results for neutrons and light isobars, 
this ratio is reduced at target rapidity with the $NL\rho\delta$ model.
Such effect strongly supports our understanding of the transport
properties of the various effective interactions. 
Pions are produced from the decay of $\Delta$ resonances formed
during the high density phase, see Fig.\ref{Fig1}(top-left). 
The $\pi^{-}$ abundance is then linked to the neutron-excess of the 
high density matter, that can form negative charged resonances, 
as detailed described in \cite{gait04}. We remind that the contribution 
of the $\delta$ 
meson leads to a more repulsive field for neutrons at supra-normal densities 
and consequently to less neutron collisions and finally to  
a smaller $\pi^{-}/\pi^{+}$ ratio. 

In conclusion, we have studied the density dependence of the isovector part of 
the nuclear $EOS$, which is still poorly known experimentally, controversially 
predicted by theory, but of great interest in extreme nuclear systems. Here we 
discuss its high density behavior that can be tested in relativistic 
heavy ion collisions.  

We have analyzed the stopping and isospin 
transparency in relativistic collisions in terms of the imbalance ratio of 
different 
particle types in projectile/target rapidity regions in mixed collisions.
We show that this observable is even sensitive to the microscopic Lorentz
structure of the symmetry term. Effective interactions with symmetry energies
not much different at $2-3~times$ the normal density $\rho_{sat}$ are
predicting large differences in the $isospin-transparencies$, depending on
the relative contribution of the various charged vector and scalar fields.
The interest is also in the fact that this appears to be a genuine
relativistic effect. Such result, combined to a weak dependence
on the in-medium $NN$ cross sections for hard nucleon collisions,
indicates the importance of imbalance ratio measurements for a
better knowledge of the nuclear $EOS$, and its $fine-structure$,
at high baryon density.

We have found moderate isospin effects on the degree of 
stopping for 
protons and neutrons, 
but important effects
in the imbalance observable 
for the ratio of tritons to ${}^{3}He$ fragments and for the ratio 
of negative to 
positive charged pions. A comparison with the few preliminary 
available data indicates 
a stiff behavior of the symmetry energy, which in our formulation 
means that the 
inclusion of $\delta$-like fields is favoured. However, more experimental 
information, 
as proposed here, would be necessary in order to achieve a clear 
definitive conclusion.



\begin{thebibliography}{00}
\bibitem{comp} We also like to remind that in any case the isospin
equilibration is not much sensitive to the stiffness of the $EOS$
in symmetric matter, as remarked in the ref.\cite{fopi3} on the
basis of $IQMD$ calculations.
\end{thebibliography}
\end{document}